\documentclass[12pt,a4paper]{article}
\usepackage{amsthm}
\usepackage{amsfonts,amsmath}
\usepackage{lscape}
\def\bea{\begin{eqnarray}}
\def\eea{\end{eqnarray}}
\def\nn{\nonumber}

\def\ket#1{\left| #1\right\rangle}
\def\bracket#1#2{\left\langle #1 | #2 \right\rangle}
\def\NN{ {\mathbb Z}_{+} }
\def\v0{ |d,r) }

\def\Z{{\mathbb Z}}
\def\half{\frac{1}{2}}
\newtheorem{lemma}{Lemma}%[section]

\newtheorem{thm}[lemma]{Theorem}

\title{On irreducible representations of the exotic conformal Galilei algebra}
\author{
Naruhiko Aizawa\footnote{ Department of Mathematics and Information Sciences, Graduate School of Science, Osaka Prefecture University,
Nakamozu Campus, Sakai, Osaka 599-8531, Japan.
aizawa@mi.s.osakafu-u.ac.jp} \ and 
Phillip S Isaac\footnote{School of Mathematics and Physics, The University of Queensland, St Lucia QLD 4072, Australia.
psi@maths.uq.edu.au }
}
\date{}

\begin{document}
\maketitle

\begin{abstract}
We investigate the representations of the exotic conformal Galilei algebra.
This is done by explicitly
constructing all singular vectors within the Verma modules, and then deducing irreducibility
of the associated highest weight quotient modules. A resulting classification of infinite
dimensional irreducible modules is presented.
\end{abstract}

\section{Introduction}

Since the proposal of a nonrelativistic analogue of the AdS/CFT correspondence \cite{Son,BaMcG}, 
conformal invariance in nonrelativistic physics 
has attracted renewed interest. 
Originally the correspondence was discussed for systems admitting the full Schr\"odinger symmetry group. 
It was recognised soon after that the correspondence was able to extend to a wider class of systems (see for example 
\cite{KLM,Taylor,Wen,BG,ADV,MT}).  
In this article we focus on the symmetry, discussed in \cite{MT}, generated by the conformal Galilei algebra with a so-called exotic 
central extension. This is an example of a non-semisimple Lie algebra whose representation theory 
has not been explored in great detail.

The exotic central extension is crucial for the holographic construction presented in \cite{MT}, although the 
obtained bulk spacetime is a ``wrong-signature'' version of AdS${}_7$. 
The same algebra also plays an important role in some models from classical mechanics with 
higher order time derivatives \cite{StZak,LSZ,LSZ2}. 
Furthermore, it is  known that there are many physical systems 
admitting extended (not necessarily conformal) Galilei symmetry with the exotic central extension 
(see for example \cite{Horva,HoMaSt} and references therein). 
We thus believe that the conformal Galilei algebra (CGA) with the exotic 
central extension is of physical importance and 
that the study of representations of the algebra 
will be useful for further investigation.

The CGA \cite{HaPle} is an extension of the Galilei algebra which generates the basic symmetry of nonrelativistic systems. 
The class of CGA we focus on is labelled by a positive half-integer $\ell $ \cite{NdelOR}. 
The smallest example $ \ell=1/2$ has the generators of conformal and scale transformations in addition to those of the Galilei 
algebra. The CGA of this case corresponds to the Schr\"odinger algebra without central extension. 
The $ \ell=1 $ algebra comprises constant accelerations and we will study a particular case of this type. 
The CGA admits two types of central extension \cite{MT,StZak,LSZ,LSZ2}. 
One exists for half-integral values of $ \ell $ and any dimension of space time. 
The other exists only in $(2+1)$ dimensions with integral values of $\ell,$ which is the
motivation for naming the central extension \textit{exotic}.

In this article we study the highest weight representations, especially the Verma modules,
over the exotic CGA in full detail. 
The classification of the irreducible modules is provided by the method similar to that
for semisimple Lie algebras. 
After presenting the structure of the CGA in section 2, we give the Verma modules in
section 3 and then proceed to construct all singular vectors and present a complete list
of infinite dimensional irreducible representations in section 4. We remark that our
presentation is similar to that of \cite{DoDoMr} for the case of the Schr\"odinger algebra. 
We also remark that the vector field realisation of the CGA is given in \cite{ADV,MT} and the 
representation of an infinite dimensional extension of CGA is studied in \cite{BaMa}.

\section{Structure of the exotic conformal Galilei algebra}

For $l\in \half \Z$, the ``spin-$l$'' class \cite{NdelOR} of conformal extensions of the
Galilei algebra is realised by the following infinitesimal action on $(d+1)$ dimensional space-time:
$$
H = \frac{\partial}{\partial t},\ 
D=-t\frac{\partial}{\partial t}-lx_i\frac{\partial}{\partial x_i},\ 
C = t^2\frac{\partial}{\partial t}+2ltx_i\frac{\partial}{\partial x_i},
$$
$$
J_{ij} = -x_i\frac{\partial}{\partial x_j}+x_j\frac{\partial}{\partial x_i},\ 
P_i^n = (-t)^n\frac{\partial}{\partial x_i},
$$
where $n=0,\ldots,2l$ and $i=1,\ldots,d$.

The algebra under consideration in this article is a conformal Galilei algebra with a
central extension. This particular central extension only exists in $(2+1)$ dimensional
spacetime with integer spin $\ell.$ Consequently this algebra is sometimes referred to as
the \textit{exotic conformal Galilei algebra} in the literature \cite{MT,LSZ,LSZ2}. 
We follow the notations and conventions of \cite{MT} studying the
case $\ell = 1$ and $d=2$.

A convenient notation introduced in \cite{MT} is to redefine the generators 
as $P_i = P_i^0, K_i = P_i^1, F_i =
P_i^2,$ for $i=1,2$, and to define a central element $\Theta$ in addition to those generators $D$,
$C$, $H$ and $J_{12}$ given above, so that we have the following non-zero commutation
relations:
$$
  \begin{array}{lll}
    [D, H] = H,  & [C, D] = C, & [C, H] = 2 D,  \\[5pt]
    [H, K_i] = -P_i, & [D, P_i] = P_i, & [C, P_i] = 2 K_i,  \\[5pt]
    [H, F_i] = -2K_i, & [D, F_i] = -F_i, & [C, K_i] = F_i,  \\[5pt]
    [J_{12}, X_1 ] = X_2, & [J_{12}, X_2] = -X_1, &
      [K_i, K_j] = \Theta \epsilon_{ij}, \quad [P_i, F_j] = -2 \Theta \epsilon_{ij},
  \end{array}
$$
where $ X_i = P_i, K_i$ or $F_i$ and $ \epsilon_{12} = -\epsilon_{21} = 1 $ is an
antisymmetric tensor. 

In order to study the representation theory of this algebra, we seek a triangular
decomposition and introduce certain linear combinations of those generators given by
$$
  X_{\pm} = X_1 \pm i X_2, \qquad J = i J_{12}, \qquad \Theta = i \Theta.
$$
The non-zero commutators in terms of these new generators are then
$$
  \begin{array}{lll}
    [ J, X_{\pm} ]  = \pm X_{\pm}, & & \\[5pt]
    [ H, K_{\pm} ] = -P_{\pm}, & [D, P_{\pm}] = P_{\pm}, & [C, P_{\pm}] = 2K_{\pm},
    \\[5pt]
    [H, F_{\pm}] = -2 K_{\pm}, & [D, F_{\pm}] = -F_{\pm}, &[C, K_{\pm}] = F_{\pm},
    \\[5pt]
    [K_+, K_-] = -2 \Theta, & [P_{\pm}, F_{\mp}] = \pm 4 \Theta. &
  \end{array}
$$
%According to the commutators involving $ D $ and $ J $ we assign the following degree to each elements:
%\bea
% & & 
%  \mbox{deg} H = (1,0), \quad \mbox{deg} C = (-1, 0), \quad \mbox{deg} P_{\pm} = (1, \pm 1), 
%  \quad
%  \mbox{deg} K_{\pm} = (0, \pm 1), 
%  \nn \\
% & & 
%  \mbox{deg} F_{\pm} = (-1, \pm 1), \quad
%  \mbox{deg} D = \mbox{deg} J = \mbox{deg} \Theta = (0,0).
%  \label{Degree}
%\eea
%Reading the integers in parenthesis from left to right, if the first non-zero number
%encountered is positive (negative), 
%then the element is referred to as positively (negatively) graded. 
%This leads us to the following definition of 
The triangular decomposition of the algebra then follows:
\bea
  & & {\mathfrak g}^+ = \{ \ H,\ P_{\pm},\  K_+ \ \} \simeq \{\ H, \ P_+,\ K_+ \ \} \oplus \{\ P_- \ \}, \nn \\[5pt]
  & & {\mathfrak g}^0= \{ \ D,\  J,\  \Theta \ \}, \nn \\[5pt]
  & & {\mathfrak g}^- = \{ \ C,\  F_{\pm},\  K_- \ \} \simeq \{ \ C, \ F_-, \ K_- \ \}
\oplus \{\ F_+ \ \}. \nn 
\eea
Note that $ {\mathfrak g}^{\pm} $ is non Abelian. Each non Abelian part is isomorphic to the Heisenberg algebra.

\section{Highest weight representations and Verma modules} 

To investigate highest weight representations of this algebra, we let $\ket{d,r}$ be a
highest weight vector such that
\bea
 & & 
  D \ket{d,r} = d \ket{d,r}, \qquad J \ket{d,r} = r \ket{d,r}, \qquad \Theta \ket{d,r} = \theta \ket{d,r},
 \nn \\
 & & 
  H \ket{d,r} = P_{\pm} \ket{d,r} = K_+ \ket{d,r} = 0.
 \nn
\eea
For fixed values of $d$ and $r$, the Verma module associated to this highest weight vector is then determined by $ V^{d,r}
= U({\mathfrak g}^-) \ket{d,r},$ where $U({\mathfrak g}^-)$ is the universal enveloping algebra of
${\mathfrak g}^-.$ Hence we are able to give a basis of $ V^{d,r} $ as
$\displaystyle{ \ket{h,k,\ell,m} = C^h K_-^k F_-^{\ell} F_+^m \ket{d,r}. } $

It is also straightforward to give the action of the generators on the basis. Here we only
give the action of generators from ${\mathfrak g}^0$ and ${\mathfrak g}^+$:
\bea
%C \ket{h,k,\ell,m} & = & \ket{h+1,k,\ell,m}, \nn \\[5pt]
%K_- \ket{h,k,\ell,m} & = & \ket{h,k+1,\ell,m} - h \ket{h-1,k,\ell+1,m}, 
%  \nn\\[5pt]
%F_- \ket{h,k,\ell,m} & = & \ket{h,k,\ell+1,m}, \nn \\[5pt]
%      F_+ \ket{h,k,\ell,m} & = & \ket{h,k,\ell,m+1}
%  \nn \\[5pt]
%  & & 
%  \nn \\
D \ket{h,k,\ell,m} & = & (d - h - \ell-m) \ket{h,k,\ell,m}, 
  \nn \\[5pt]
J \ket{h,k,\ell,m} & = & (r -k - \ell+m) \ket{h,k,\ell,m},
  \nn \\
  & & 
  \nn \\[5pt]
H \ket{h,k,\ell,m} & = & -2\ell \ket{h,k+1,\ell-1,m} + 4km \theta \ket{h,k-1,\ell,m-1}
  \nn \\
  & & \hspace{2cm}
      + h (2\ell + 2m + h-2d-1) \ket{h-1,k,\ell,m},
  \nn \\[5pt]
K_+ \ket{h,k,\ell,m} & = & -2k \theta \ket{h,k-1,\ell,m} - h \ket{h-1,k,\ell,m+1},
  \nn \\[5pt]
P_+ \ket{h,k,\ell,m} & = & 4 \ell \theta \ket{h,k,\ell-1,m} + 4 h k \theta \ket{h-1,k-1,\ell,m}
  \nn \\
  & & \hspace{2cm}
     + h (h-1) \ket{h-2,k,\ell,m+1},
  \nn \\[5pt]
P_- \ket{h,k,\ell,m}  & = & -4m \theta \ket{h,k,\ell,m-1} -2h \ket{h-1,k+1,\ell,m} 
  \nn \\
  & & \hspace{2cm}
     +h (h-1) \ket{h-2,k,\ell+1,m}.
  \nn 
\eea
We note that $D$ and $J$ are diagonal on this basis. We set
\[
   p = h + \ell + m \geq 0 , \qquad q = k + \ell - m \quad \in {\mathbb Z}
\]
Then we can express the action of $D$ and $J$ on the basis as  
$$
  D \ket{h,k,\ell,m} = (d-p) \ket{h,k,\ell,m}, \qquad J \ket{h,k,\ell,m} = (r-q) \ket{h,k,\ell,m}.
$$
It is then clear that the Verma module has a weight decomposition for fixed values of $ p $ and $ q:$ 
\begin{equation}
  V^{d,r} = \bigoplus V^{d,r}_{d-p, r-q}  \label{WeghtDecomp}
\end{equation}

\section{Singular vectors and irreducible modules}

A singular vector is a homogeneous element with respect to the decomposition
(\ref{WeghtDecomp}). Its general form is
$$
\ket{v} = \sum_{\ell,m} a_{\ell,m} \ket{p-\ell-m, q-\ell+m, \ell,m}.
$$
In other words, a singular vector 
is a linear combination of a subset of the basis of the Verma module corresponding to fixed values of $ p $ and $q. $ 
We have, by definition, 
$h = p - \ell-m \geq 0$ and $k = q -\ell + m \geq 0$
from which it follows that
\begin{equation}
  0 \leq \ell+ m \leq p, \label{lm-range}
\end{equation}
and
\begin{equation}
  \ell \leq m + q \quad (0 \leq q), \qquad\quad \ell + |q| \leq m.
  \label{m-range}
\end{equation}
The most general form of the singular vector for given values of $p$ and $q$ can then be
expressed as
\begin{equation}
  \ket{v} = \sum_{\ell=0}^{\lfloor \frac{p+q}{2} \rfloor} \sum_{m=\ell-q}^{p-\ell} 
              a_{\ell,m} \ket{p-\ell-m,q-\ell+m,\ell,m},
  \label{SVgen}
\end{equation}
where we understand that $ a_{\ell,m} = 0 $ for the pair $(\ell,m)$ not 
satisfying (\ref{lm-range}) or (\ref{m-range}). 
The condition for $ \ket{v} $ being a singular vector is given by
$H \ket{v} = P_{\pm} \ket{v} = K_+ \ket{v} = 0.$ 

In order to fully understand if we can determine the coefficients for the singular vector at a specific level, we
need to consider three different cases, namely $q>0$, $q=0$ and $q<0$.

\noindent
\underline{Case $q>0$}:

\noindent
In this case one can rewrite (\ref{SVgen}) as follows:
\begin{equation}
   \ket{v} =  \sum_{m=0}^p \sum_{\ell=0}^{\min\{p-m,q+m\}} 
              a_{\ell,m} \ket{p-\ell-m,q-\ell+m,\ell,m}.
    \label{SValt1}
\end{equation}
It is not difficult to see that the action of $ K_+ $ on (\ref{SValt1}) is calculated as 
follows:
\bea
\hspace{-1cm}  K_+ \ket{v} &=& - \sum_{\ell=0}^{\min\{p, q-1 \} } 2 \theta (q-\ell)\, a_{\ell,0}\, 
                \ket{p-\ell,q-1-\ell,\ell,0}
   \nn \\
   & & - \sum_{m=1}^p \sum_{\ell=0}^{\min\{p-m,q-1+m  \} } 
   \{ \ 2\theta (q-\ell+m)\, a_{\ell,m} + (p-\ell-m+1)\, a_{\ell,m-1} \ \}
   \nn \\
   & & \hspace{4cm} \times \ket{p-\ell-m,q-\ell+m-1,\ell,m}.
   \nn
\eea
The condition $ K_+ \ket{v} = 0 $ yields one recurrence relation 
\begin{equation}
  2\theta (q-\ell+m)\, a_{\ell,m} + (p-\ell-m+1)\, a_{\ell,m-1} = 0,
  \label{recKplus} 
\end{equation}
for $ 1 \leq m \leq p, \ 0 \leq \ell \leq \min\{p-m,q-1+m  \} $ 
with initial condition
%% We note that this restriction on $ \ell, m $ is equivalent to 
%% $
%%  0 \leq \ell \leq \lfloor \frac{p+q}{2} \rfloor, \ \max\{\ell-q+1,1\} \leq m \leq p-\ell. 
%% $
\begin{equation}
  a_{\ell,0} = 0, \quad \quad 0 \leq \ell \leq \min\{p,q-1\}.
  \label{a-ell-0}
\end{equation}
The recurrence relation (\ref{recKplus}) is solved to give
\begin{equation}
  a_{\ell,m} = \left( -\frac{1}{2\theta} \right)^m 
    \frac{ (p-\ell)! (q-\ell)! }{ (p-\ell-m)! (q-\ell+m)! } 
    a_{\ell,0}.
  \label{alm-Kplus}
\end{equation}
From (\ref{a-ell-0}) and (\ref{alm-Kplus}) one can see the following facts:
\begin{enumerate}
  \renewcommand{\labelenumi}{(\roman{enumi})}
  \item if $ p < q $ then $ a_{\ell,m} = 0 $ for all possible pairs of $ (\ell,m). $ 
  Thus $ \ket{v} = 0. $
  \item if $ 0 < q \leq p $ then 
     $ a_{\ell,m} = 0 $ for $ 0 \leq \ell \leq q-1. $  
\end{enumerate}
We study the case (ii) further. From (\ref{SVgen}) and (\ref{alm-Kplus}) we have
$$
  \ket{v} = \sum_{\ell=q}^{\lfloor \frac{p+q}{2} \rfloor} \sum_{m=\ell-q}^{p-\ell} 
    \left( -\frac{1}{2\theta} \right)^m 
    \frac{ (p-\ell)! (q-\ell)! }{ (p-\ell-m)! (q-\ell+m)! }\, a_{\ell,0}
     \ket{p-\ell-m,q-\ell+m,\ell,m}.
$$
We define
$$
  \ket{v^{\ell}} = \sum_{m=\ell-q}^{p-\ell} \left( -\frac{1}{2\theta} \right)^m 
  \frac{ 1 }{ (p-\ell-m)! (q-\ell+m)! } \ket{p-\ell-m,q-\ell+m,\ell,m}.
$$
Then we have
$\displaystyle{\ket{v} = \sum_{\ell=q}^{\lfloor \frac{p+q}{2} \rfloor} \alpha_{\ell} \ket{v^{\ell}}, }$
and $ \{ \ \ket{v^{\ell}} \ | \ q \leq \ell \leq \lfloor \frac{p+q}{2} \rfloor \ \} $ is a 
set of linearly independent vectors in the kernel of $ K_+, $ i.e., 
$ K_+ \ket{v^{\ell}} = 0. $ 
We now calculate the action of $ P_+ $ on $ \ket{v^{\ell}}. $ 
\begin{eqnarray*}
\hspace{-1cm} P_+  \ket{v^{\ell}} &=& 
\sum_{m=\ell-q}^{p-\ell} \left( -\frac{1}{2\theta} \right)^m 
\frac{ 1 }{ (p-\ell-m)! (q-\ell+m)! } 
\\
&&\quad\times 
\{ \ 4 \ell \theta \ket{p-\ell-m,q-\ell+m,\ell-1,m} 
\\
& & 
+ 4 (p-\ell-m) (q-\ell+m) \theta \ket{p-\ell-m-1,q-\ell+m-1,\ell,m}
\\
& & 
+ (p-\ell-m) (p-\ell-m-1) \ket{p-\ell-m-2, q-\ell+m,\ell,m+1} \ \}.
\end{eqnarray*}
Looking at the highest values of $m$ the vector $ \ket{0,p+q-2\ell,\ell-1,p-\ell} $ 
on the RHS has nonvanishing coefficients for all possible values of $ \ell.$ 
It follow that the condition $ P_+ \ket{v} = 0 $ implies $ \alpha_{\ell} = 0. $ 
We thus have $ \ket{v} = 0. $ 

We therefore have shown that there are no singular vectors for $ q > 0. $

\noindent
\underline{Case $q=0$:}

\noindent
In this case we have the recurrence relation (\ref{recKplus}) from the previous case but no initial conditions. 
We solve (\ref{recKplus}) to have
\begin{equation}
  a_{\ell,m} = \left( -\frac{1}{2\theta} \right)^{m-\ell} 
  \frac{ (p-2\ell)! }{ (p-\ell-m)! (m-\ell)! } a_{\ell,\ell}. 
  \label{q0alm}
\end{equation}
The condition $ P_-v \ket{v} = 0 $ yields the relation
$$
  -4(m+1) \theta a_{\ell,m+1} - 2(p-\ell-m) a_{\ell,m} + (p-\ell-m+1) (p-\ell-m) a_{\ell-1,m} = 0.
$$
Substitution of (\ref{q0alm}) into this yields the recurrence relation
\[
  4 \ell \theta a_{\ell,\ell} - (p-2\ell+2) (p-2\ell+1) a_{\ell-1,\ell-1} = 0.
\]
This is solved to give the expression
\begin{equation}
  a_{\ell,\ell} = \left(\frac{1}{4 \theta} \right)^{\ell}  
  \frac{p!}{ \ell ! (p-2\ell)! } a_{0,0}.
  \label{q0a00}
\end{equation}
Substitution of (\ref{q0a00}) into (\ref{q0alm}) determines the coefficients as follows
\begin{equation}
  a_{\ell,\ell} = \left(-\frac{1}{2} \right)^{m+\ell} \frac{1}{\theta^m} 
  \frac{p!}{ \ell ! (m-\ell)! (p-\ell-m)! } a_{0,0}.
  \label{q0coeff}
\end{equation}
The condition $ P_+ \ket{v} = 0 $ yields another recurrence relation
$$
  4 (\ell+1) \theta a_{\ell+1,m} + 4 (p-\ell-m) (m-\ell) \theta a_{\ell,m} 
  + (p-\ell-m+1)(p-\ell-m) a_{\ell,m-1} = 0.
$$
It is easy to verify that (\ref{q0coeff}) satisfies this relation. 
Next we look at the condition $ H \ket{v} = 0. $ 
It gives the recurrence relation containing $d$
$$
  -2(\ell+1) a_{\ell+1,m} + (p-\ell-m) (p+\ell+m-2d-1) a_{\ell,m} 
  + 4(m-\ell+1) (m+1) \theta a_{\ell,m+1} = 0.
$$
Substitution of (\ref{q0coeff}) into the left hand side gives the expression
\[
   \left( - \frac{1}{2} \right)^{m+\ell} \frac{1}{\theta^m} 
   \frac{ p! a_{0,0} }{ \ell ! (m-\ell)! (p-\ell-m-1)! } 
   (p-2d-3).
\]
Setting this equal to zero we obtain the condition for $d: $  
\begin{equation}
  p=2d + 3 \ \in \NN  \label{q0d}
\end{equation}
In summary there exist one singular vector in $ V^{d,r} $ if $d$ satisfies the condition (\ref{q0d})  
and it is given by (up to overall factor)
$$
\ket{v_s} = \sum_{m=0}^p \sum_{\ell=0}^{\mbox{min}\{m, p-m\}} 
    \left( -\frac{1}{2} \right)^{m+\ell} \frac{1}{\theta^m} 
    \frac{ p! }{ \ell ! (m-\ell)!  (p-\ell-m)! } 
    C^{p-\ell-m} K_-^{m-\ell} F_-^{\ell} F_+^m \ket{d,r}.
$$
It is then possible to show by induction that the singular vector has the closed form
$$
  \ket{v_s} = (2\theta C - K_- F_+)^p \ket{d,r}.  
$$

\noindent
\underline{Case $q<0$:}

\noindent
In this case the most general form (\ref{SVgen}) of the singular vector 
yields 
$$
  \ket{v} = \sum_{\ell=0}^{\lfloor \frac{p-|q|}{2} \rfloor} \sum_{m=\ell+|q|}^{p-\ell} 
  a_{\ell,m} \ket{p-\ell-m,m-\ell-|q|,\ell,m}.
$$
The condition $ K_+ \ket{v} = 0 $ gives us the recurrence relation
\begin{equation}
  2 (m-\ell-|q|+1) \theta a_{\ell,m+1} + (p-\ell-m+) a_{\ell,m} = 0,
  \label{recKplus2}
\end{equation}
for $ 0 \leq \ell \leq {\lfloor \frac{p-|q|}{2} \rfloor}, \ \ell+|q| \leq m \leq p-\ell-1. $ 
On the other hand, the condition $ P_- \ket{v} = 0 $ yields one initial condition and 
recurrence relation, since
\bea
   & &\hspace{-2.5cm} P_- \ket{v} =
    - \left( 
         \sum_{m=|q|-1}^{p-1} 4 (m+1) \theta a_{0,m+1} + \sum_{m=|q|}^{p-1} 2 (p-m) a_{0,m} 
      \right) 
  \nn\\
  & & \hspace{4cm} \times
      \ket{p-m-1,m-|q|+1,0,m}
  \nn \\
  & &\hspace{-2.5cm}  -\sum_{\ell=1}^{\lfloor \frac{p-|q|}{2} \rfloor} \{ \ 4 (\ell+|q|) \theta a_{\ell,\ell+|q|}
     - (p-|q|-2\ell+2) (p-|q|-2\ell+1) a_{\ell-1,\ell-1+|q|} \ \}
  \nn \\
  & & \hspace{4cm} \times \ket{p-|q|-2\ell,0,\ell,\ell-1+|q|}
  \nn \\
  & &\hspace{-2.5cm}  - \sum_{\ell=1}^{\lfloor \frac{p-|q|}{2} \rfloor} \sum_{m=\ell+|q|}^{p-\ell-1} 
      \{ \ 
        4(m+1) \theta a_{\ell,m+1} + 2(p-\ell-m) a_{\ell,m}
        - (p-\ell-m+1) (p-\ell-m) a_{\ell-1,m} \ \}
  \nn \\
  & & \hspace{4cm} \times \ket{p-\ell-m-1,m-\ell-|q|+1,\ell,m}.
  \label{PminusAction}
\eea
Now we only inspect the important relations. Looking at $ m = |q|-1 $ we have the initial condition 
$ a_{0,|q|} = 0. $ 
The first line in (\ref{PminusAction}) gives the recurrence relation
\begin{equation}
   2(m+1) \theta a_{0,m+1} + (p-m) a_{0,m} = 0,
   \label{RecRel1}
\end{equation}
for $ |q| \leq m \leq p-1. $ Recursive use of (\ref{RecRel1}) with $ a_{0,|q|} = 0 $ shows 
that $ a_{0,m} = 0 $ for $ 0 \leq m \leq p. $ 
From the last line of (\ref{PminusAction}) we obtain 
the recurrence relation
\begin{equation}
        4(m+1) \theta a_{\ell,m+1} + 2(p-\ell-m) a_{\ell,m}
        - (p-\ell-m+1) (p-\ell-m) a_{\ell-1,m} = 0.
        \label{recPminus}
\end{equation}
Substitution of (\ref{recKplus2}) into (\ref{recPminus}) gives the relation
\begin{equation}
   2(\ell+|q|) a_{\ell,m+1} + (p-\ell-m) (m-\ell-|q|+2) a_{\ell-1,m+1} = 0.
   \label{rec-q-negative}
\end{equation}
One can show by the relation (\ref{rec-q-negative}) that $ a_{\ell,m} = 0 $ for 
all possible pairs of $ (\ell, m). $ 
We thus conclude there are no singular vectors for $ q < 0. $ 

The results of the preceding discussion can be nicely summarised in the following theorem. 
\begin{thm} \label{thm1}
$ V^{d,r} $ has precisely one singular vector iff $ 2d + 3 \ \in \NN $ and it is given by 
\begin{equation}
  \ket{v_s} = (2\theta C - K_- F_+)^{2d+3} \ket{d,r}.\label{q0SVclosed}
\end{equation}  
\end{thm}

As a remark, we note that we are able to define a bilinear form $(\ ,\ )$ on $V^{d,r}$, following
Shapovalov \cite{Shap}, such that
\bea
  & & \bracket{d,r}{d,r} \equiv (\; \ket{d,r}, \ket{d,r} \;) = 1,
  \nn \\[5pt]
  & & ( A \ket{d,r}, B \ket{d,r}) = (\ket{d,r}, \omega(A) B \ket{d,r}),
  \label{Bilinear1}
\eea
where $\omega$ is the involutive algebra anti-automorphism defined by
$$
\omega(D) = D,\ \omega(J) = J,\ \omega(\Theta) = \Theta,\
\omega(C) = H,\ \omega(K_+) = K_-,\  \omega(P_{\pm}) = F_{\mp}.
$$
%\begin{equation}
%  \begin{array}{lll}
%    \omega(D) = D, & \omega(J) = J, & \omega(\Theta) = \Theta,
%    \\[5pt]
%    \omega(C) = H, &  \omega(K_+) = K_- &  \omega(P_{\pm}) = F_{\mp}.   
%  \end{array}
%  \label{omega1}
%\end{equation}
%Let $ \ket{d_i, r_i} $ be a state with the weight $ (d_i, r_i). $ 
%Then
%\begin{equation}
%   \bracket{d_i,r_i}{d_j,r_j} = 0, \qquad \mbox{if} \ d_i \neq d_j, \ \mbox{or} \ r_i \neq r_j
%   \label{orthogonality}
%\end{equation}
%since 
%$$
%d_i \bracket{d_i,r_i}{d_j,r_j} = (D\ket{d_i,r_i}, \ket{d_j,r_j}) = 
%  ( \ket{d_i,r_i}, \omega(D) \ket{d_j,r_j} ) = d_j \bracket{d_i,r_i}{d_j,r_j}
%$$
It is straightforward to see that the singular vectors are orthogonal to all vectors in $
V^{d,r} $ with respect to the bilinear form (\ref{Bilinear1}).

The Verma modules without the singular vectors have no invariant 
submodules of highest weight type. 
In the remainder of the paper, we therefore study $ V^{d,r}$ with $ 2d+3 \in \NN.$ 

Let $ {\cal I}^d = U({\mathfrak g}^-) \ket{v_s} $  where $ \ket{v_s} $ 
is the singular vector (\ref{q0SVclosed}). Then $ {\cal I}^d $ is an 
invariant submodule and we consider the quotient module 
$ \tilde{V}^{d,r} = V^{d,r}/{\cal I}^d. $ 
The highest weight vector in $ \tilde{V}^{d,r}, $ denoted by $\v0,$  is defined by 
\bea
  & & D \v0 = d \v0, \qquad J \v0 = r \v0, \qquad \Theta \v0 = \theta \v0, 
    \nn \\[5pt]
  & & H \v0 = P_{\pm} \v0 = K_+ \v0 = (2\theta C - K_- F_+)^{2d+3} \v0 = 0.
    \label{HWVinQuotient}
\eea
The basis of $ \tilde{V}^{d,r} $ has the form of 
$ |h,k,\ell,m) = C^{h} K_-^{k} F_-^{\ell} F_+^m \v0. $
However because of (\ref{HWVinQuotient}) the vector $ C^{2d+3} \v0 $ is not independent. 
%As an example, we give the relation for $ 2d+3 = 1, 2:$
%\bea
%  & & C \v0 = \frac{1}{2\theta} K_- F_+ \v0, 
%  \nn \\
%  & & C^2 \v0 = 
%      \left(
%         \frac{1}{\theta} C K_- F_+  - \frac{1}{2\theta} F_- F_+ - \frac{1}{4\theta^2} K_-^2 F_+^2  
%      \right) \v0.
%  \nn
%\eea 
Next we look for the singular vectors in $ \tilde{V}^{d,r} $ by the same procedure as
the earlier discussion in this section. 
The singular vector has the form of 
\[
  | v) = \sum_{\ell,m} \alpha_{\ell,m}\, |p-\ell-m,m-\ell,\ell,m). 
\]
We note that $ \alpha_{0,0} $ is missing from the summation. 
As seen from the previous discussion, nonvanishing $ \alpha_{0,0} $  is 
crucial for the existence of the singular vector. 
We thus conclude that there is no singular vector in $ \tilde{V}^{d,r}, $ and hence
arrive at the main result of this article.

\begin{thm}
All irreducible highest weight modules over the exotic conformal Galilei algebra are listed as follows:
\begin{enumerate}
    \renewcommand{\labelenumi}{(\roman{enumi})}
    \item The Verma module $ V^{d,r}$ for $ 2d + 3 \notin \NN $
    \item The quotient module $ \tilde{V}^{d,r} \subseteq  V^{d,r} $ for $ 2d + 3 \in \NN$    
\end{enumerate}
where $ d, r \in {\mathbb R}. $ All irreducible modules given are infinite dimensional. 
\end{thm}

\section{Concluding remarks}

In this article we have determined all irreducible highest weight modules of the conformal
Galilei algebra with exotic central extension. This algebra has attracted attention
recently due to its application to ``exotic'' physical systems \cite{MT,StZak,LSZ,LSZ2,Horva,HoMaSt}. It was suggested in \cite{MT} that the
results of their paper relating to the exotic conformal Galilei algebra may be beneficial in the further
study of such systems, and our view is that understanding the representation
theory of the symmetry algebra of these systems will enable future developments. The
current article is an important step in developing this stratagem.

An immediate venture would be to investigate hierarchies of partial differential equations
(PDEs) associated with the singular vectors obtained in Theorem \ref{thm1} of the current paper.
Such PDEs arise via the vector field realisation of the generators in terms of
differential operators, and their action on an appropriate function space related to the
Verma modules (see \cite{Kost,Dobr} for the semisimple case). Indeed, an analogue of this was done
explicitly in \cite{DoDoMr} for the Schr\"odinger algebra in (1+1) dimensions, where a
hierarchy of generalised heat/Schr\"odinger equations was found.

Finally, it would be desirable to provide a generalised formalism for describing the
representations of families of non-semisimple Lie algebras which contain structures such as the
Schr\"odinger algebras, the conformal Galilei algebras and their central extensions.
Establishing such a comprehensive framework that recovers our results and those of
\cite{DoDoMr} (amongst others) would be a worthwhile program that could result in a broad
class of (nonrelativistic) physical systems.

\section*{Acknowledgements}

The majority of this work was done while NA was visiting the School of Mathematics
and Physics at the University of Queensland as a Raybould Fellow. We appreciate the
support of this program.

\end{document}